\documentclass[twocolumn,showpacs,superscriptaddress,amsmath,amssymb,prl]{revtex4}

\usepackage{graphicx}%
\usepackage{dcolumn}%

\begin{document}

\title{Unadulterated spectral function of low energy quasiparticles: Bi-2212, nodal direction}

\author{D. V. Evtushinsky}
\affiliation{Institute of Metal Physics of National Academy of Sciences of Ukraine, 03142 Kyiv, Ukraine}

\author{A. A. Kordyuk}
\affiliation{Institute of Metal Physics of National Academy of Sciences of Ukraine, 03142 Kyiv, Ukraine}
\affiliation{IFW Dresden, P.O. Box 270116, D-01171 Dresden, Germany}

\author{S. V. Borisenko}
\author{V.\;B.\;Zabolotnyy}
\author{M. Knupfer}
\author{J. Fink}
\author{B.\;B\"uchner}
\affiliation{IFW Dresden, P.O. Box 270116, D-01171 Dresden, Germany}

\author{\\A. V. Pan}
\affiliation{ISEM, University of Wollongong, Wollongong, NSW 2522, Australia}

\author{A. Erb}
\affiliation{Walther-Mei\ss ner-Institut, Bayerische Akademie der Wissenschaften, 85748 Garching, Germany}

\author{C. T. Lin}
\affiliation{Max-Planck Institut f\"ur Festk\"orperforschung, 70569 Stuttgart, Germany}

\author{H. Berger}
\affiliation{Institut de Physique de la Mati\'ere Complexe, EPFL, 1015 Lausanne, Switzerland}

\date{May 1, 2005}%

\begin{abstract}
Fitting the momentum distribution photoemission spectra to the Voigt profile appears to be a robust procedure to purify the interaction effects from the experimental resolution. In application to Bi-2212 high-$T_c$ cuprates, the procedure reveals the \textit{true} scattering rate at low binding energies and temperatures, and, consequently, the true value of the elastic scattering. Reaching the minimal value $\sim$\;16\;meV, the elastic scattering does not reveal a systematic dependence on doping level, but is rather sensitive to impurity concentration, and can be explained by the forward scattering on out-of-plane impurities. The inelastic scattering is found to form well-defined quasiparticles with the scattering rate $\sim \omega^2$ and $\sim \omega^3$, above and below $T_c$, respectively. The observed $\omega - T$ asymmetry of the scattering rate is also discussed.
\end{abstract}

\pacs{74.25.Jb, 74.72.Hs, 79.60.-i, 71.15.Mb}%

\preprint{}

\maketitle

A century ago, in time of the renowned Kamerlingh Onnes's experiments, the search for actual behavior of electrical conductivity in pure metals at low temperature has led to discovery of superconductivity \cite{Buckel}. Today, the similar problem, the low temperature/energy behavior of normal electrical conductivity, is topical again, now for high-$T_c$ cuprates (HTSC). The quantity of special interest is a quasiparticle scattering rate $1/\tau$ \cite{QP} or, more precisely, the single particle self-energy, $\Sigma = \Sigma' + i\Sigma''$ \cite{Hufner}, the real part of which can be associated with the mass renormalization and the imaginary part is proportional to the scattering rate or, in the simplest Drude model, to the normal state resistivity. The true lowest value of $\Sigma''$ (taken at the Fermi level) is important to know in order to reconcile the parameters of quasiparticle spectrum with transport measurements, but its asymptotic behavior, i.e.~$\Sigma''(\omega,T)$ at low energy and temperature, is vital to judge whether the quasiparticle approach is applicable at all to describe the electronic properties of HTSC.

The self-energy function, $\Sigma(\omega,T)$, is closely related to (and can be derived from) the quasiparticle spectrum, presented by the quasiparticle spectral function $A(\mathbf{k},\omega,T)$, which, in turn, can be mapped accurately in the whole Brillouin zone (BZ) by modern angle resolved photoemission spectroscopy (ARPES) \cite{ARPES}. For the nodal direction, along which the superconducting $d$-wave gap function has a node (i.e.~changes sign), all the interactions which form the nodal quasiparticles are encapsulated in $\Sigma$, both parts of which can be confidently determined from ARPES spectra in the range about 0.3 eV below the Fermi level \cite{KordyukPRB2005}. Nevertheless, the detailed behavior of $A(\omega, T)$ and $\Sigma(\omega,T)$ in the very vicinity to $E_F$ remains puzzling. The experimental resolution, which can be safely neglected for higher binding energy, plays a crucial role here. Roughly, the photocurrent intensity can be well approximated by a convolution of the spectral function, multiplied by the Fermi-function $f(\omega)$, and overall experimental resolution \cite{BorisenkoPRB2001}: 
\begin{eqnarray}\label{I}
I(\mathbf{k},\omega) \propto A(\mathbf{k},\omega) f(\omega) \otimes R(\mathbf{k},\omega).
\end{eqnarray}
The latter consists of two essential components, the response function of the analyser, $R_A(\mathbf{k},\omega)$, and one which accounts for inhomogeneities of sample surface, $R_S(\mathbf{k},\omega)$: $R = R_A \otimes R_S$. While the analyser response function is fixed and can be measured independently, the surface inhomogeneities (mechanical, chemical or in charge distribution) result in a systematic error which is difficult to account for. This seemingly technical problem has rather strong fundamental impact, setting an unavoidable limit for $\Sigma''$ estimation accuracy. In this paper we show that using a simple line shape analysis---namely a Voigt fitting procedure---one can purify the intrinsic interaction effects from the extrinsic influence of the experimental setup and, therefore, uncover true parameters of low energy part of quasiparticle spectrum. Applying the procedure to the nodal photoemission spectra from Bi$_2$Sr$_2$CaCu$_2$O$_{8+\delta}$ (Bi-2212), we determine true values of the impurity scattering, as well as the energy and temperature dependence of the scattering rate in close vicinity to the Fermi level.

We analyze the spectra from pure Bi-2212, lead-doped superstructure free Bi(Pb)-2212 \cite{BorisenkoPRB2001, KordyukPRL2004, KordyukPRB2004, KordyukPRB2005}, as well as Bi-2212 doped with Zn and Ni \cite{Zn}. Here we focus on the spectra measured along the nodal ($\pi,\pi$) direction where the 5$\times$1 superstructure is well resolved \cite{BorisenkoPRL2000} and at 27\;eV excitation energy at which the contribution from the bonding band is essentially suppressed \cite{KordyukPRB2004}. The experimental details can be found elsewhere \cite{BorisenkoPRB2001, KordyukPRL2004, KordyukPRB2004, KordyukPRB2005, Zn}.

\begin{figure}[!t]
\includegraphics[width=8.8cm]{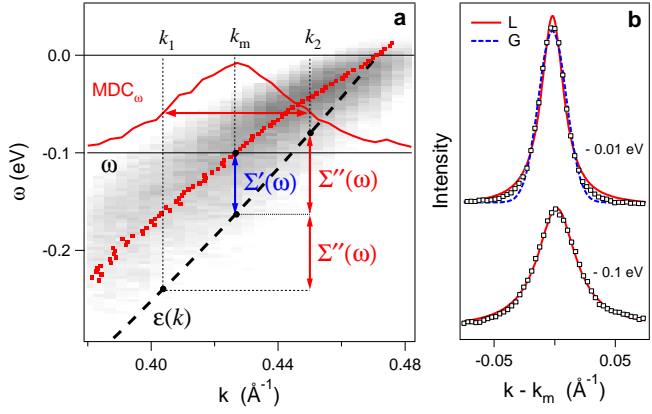}%
\caption{\label{MDCs} (a) Nodal spectra parameters: Bare band dispersion (dashed line) and renormalized dispersion (points) on top of the spectral weight of interacting electrons (``ARPES image''). Solid line represents a single MDC at $\omega$, arrows indicate its FWHM and the self-energy parts derived for given MDC. (b) Two MDC's taken at $-0.1$\;eV and $-0.01$\;eV, and their fits to Lorentzian (solid lines) and Gaussian (dashed line) functions.}
\end{figure}

Fig.\;\ref{MDCs}a introduces the essentials of the nodal spectra analysis. The blurred region represents the ``ARPES image"---the photocurrent intensity, $I(k,\omega)$, which, over the occupied states, can be well approximated by the quasiparticle spectral function
\begin{eqnarray}\label{A}
A(k, \omega) = -\frac{1}{\pi}\frac{\Sigma''(\omega)}{(\omega - \varepsilon(k) - \Sigma'(\omega))^2 + \Sigma''(\omega)^2},
\end{eqnarray}
where $\varepsilon(k)$ is the bare band dispersion along the nodal direction. Since the ARPES image became a unit of information in modern photoemission, the advantages of the analysis of the ARPES spectra in terms of the momentum distribution curves, $\mathrm{MDC} \equiv I(k,\omega=\mathrm{const})$, had been immediately realized \cite{VallaSci1999}. The main advantage comes from the fact that $A(k)$ has a simple Lorentzian lineshape as long as momentum dependence of the self-energy and bare Fermi velocity, $v_F = d\epsilon(k)/dk$, can be neglected \cite{RanderiaPRB2004}. The latter, as well as Eq.(\ref{I}), has been shown to be valid for the nodal direction of Bi-2212 up to 0.3\;eV binding energy by means of Kramers-Kronig self-consistency of the self-energy parts \cite{KordyukPRB2005}. The relations of bare dispersion and self-energy parts with the parameters of an MDC at given $\omega$ are shown in Fig.\;\ref{MDCs}a, though, in this paper we focus in $\Sigma''$, which, at low binding energy, is simply proportional to the MDC width (half width at half maximum): $\Sigma'' = v_F W$.

The analysis presented in Ref.\;\onlinecite{KordyukPRB2005} also allows to estimate contribution of the experimental resolution, although, due to a number of parameters involved, such an estimate is not very precise. Nevertheless, the close similarity between $I(k,\omega)$ and $A(k,\omega)$ at higher binding energies suggests the way to recover true quasiparticle spectrum also in close vicinity to $E_F$. It was noticed \cite{Details} that, when approaching the Fermi level, the lineshape of MDC measured along the nodal direction evolves from almost ideal Lorentzian to more Gaussian-like, which, evidently, is a result of convolution of the photoemission signal with the total response function of the setup. Fig.\;\ref{MDCs} illustrates such a Lorentzian to Gaussian crossover: While the MDC taken at $\omega = -0.1$\;eV is almost perfect Lorentzian, the MDC measured closer to the Fermi level at $\omega = -0.01$\;eV can be fitted to neither Lorentzian nor Gaussian but to a convolution of these two---the Voigt profile. 

\begin{figure}[!t]
\includegraphics[width=6cm]{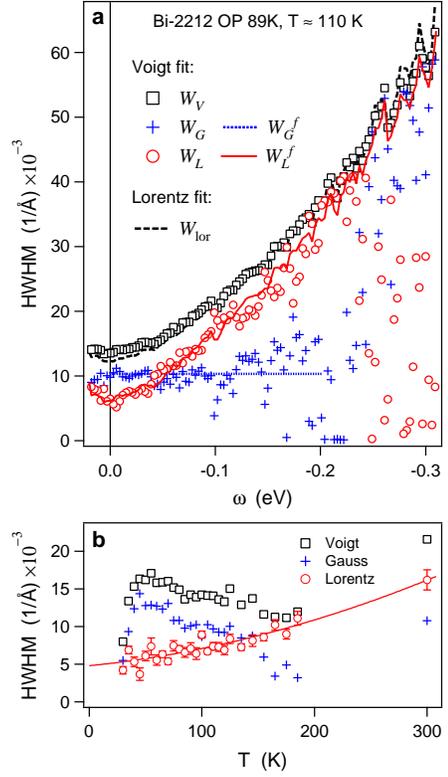}%
\caption{\label{Fig2} Constituents of MDC width: (a) energy dependence of MDC width presented by half width at half maximum (HWHM) of its fit to the Voight profile, $W_V$, ($W_{\mathrm{lor}}$ is the result of simple Lorentzian fit), and its Lorentzian, $W_L$, and Gaussian, $W_G$, constituents; (b) variation of the above mentioned parameters (at $\omega = 0$) with temperature.}
\end{figure}

\begin{figure*}[!t]
\includegraphics[width=5.4cm]{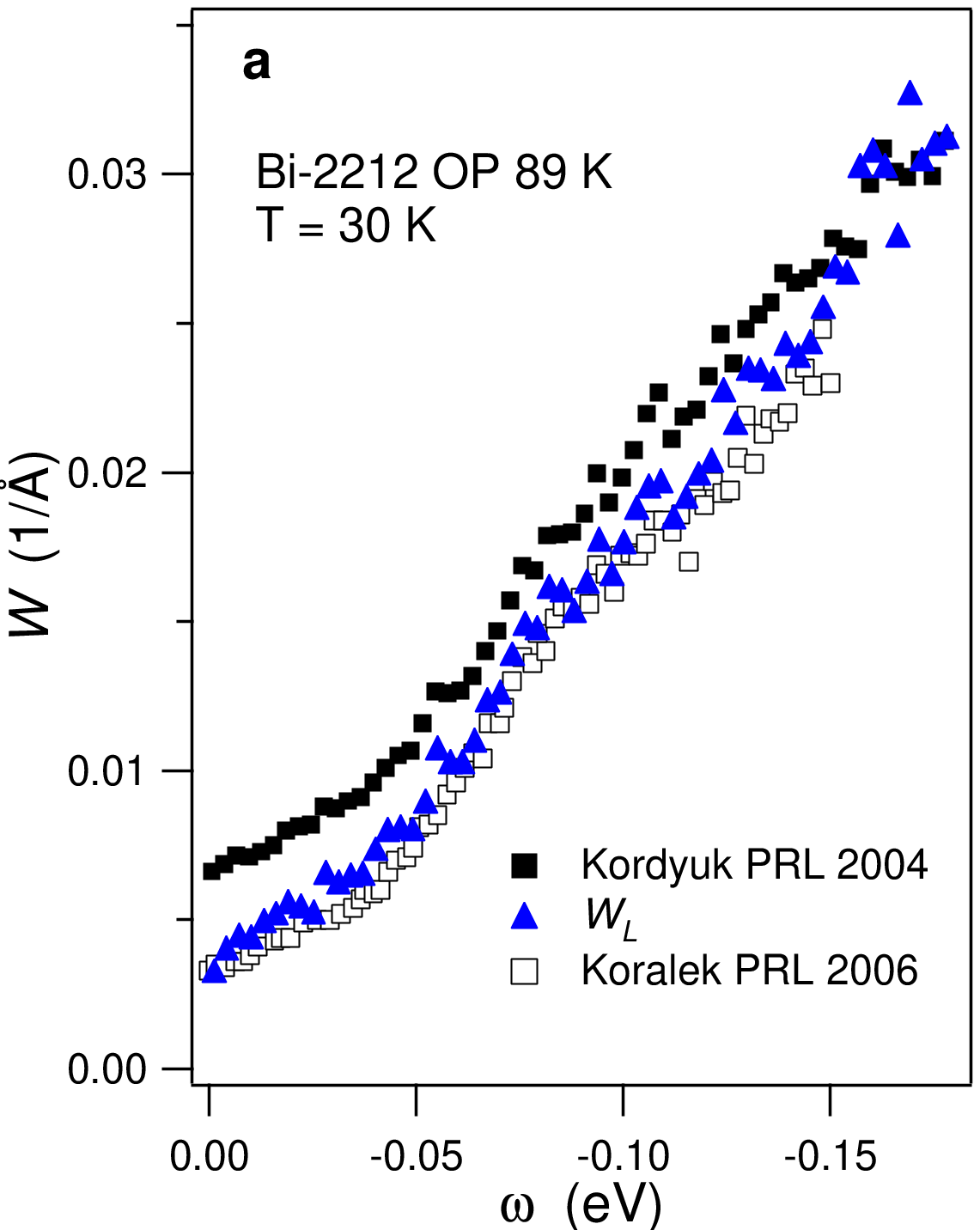}\;%
\includegraphics[width=5.4cm]{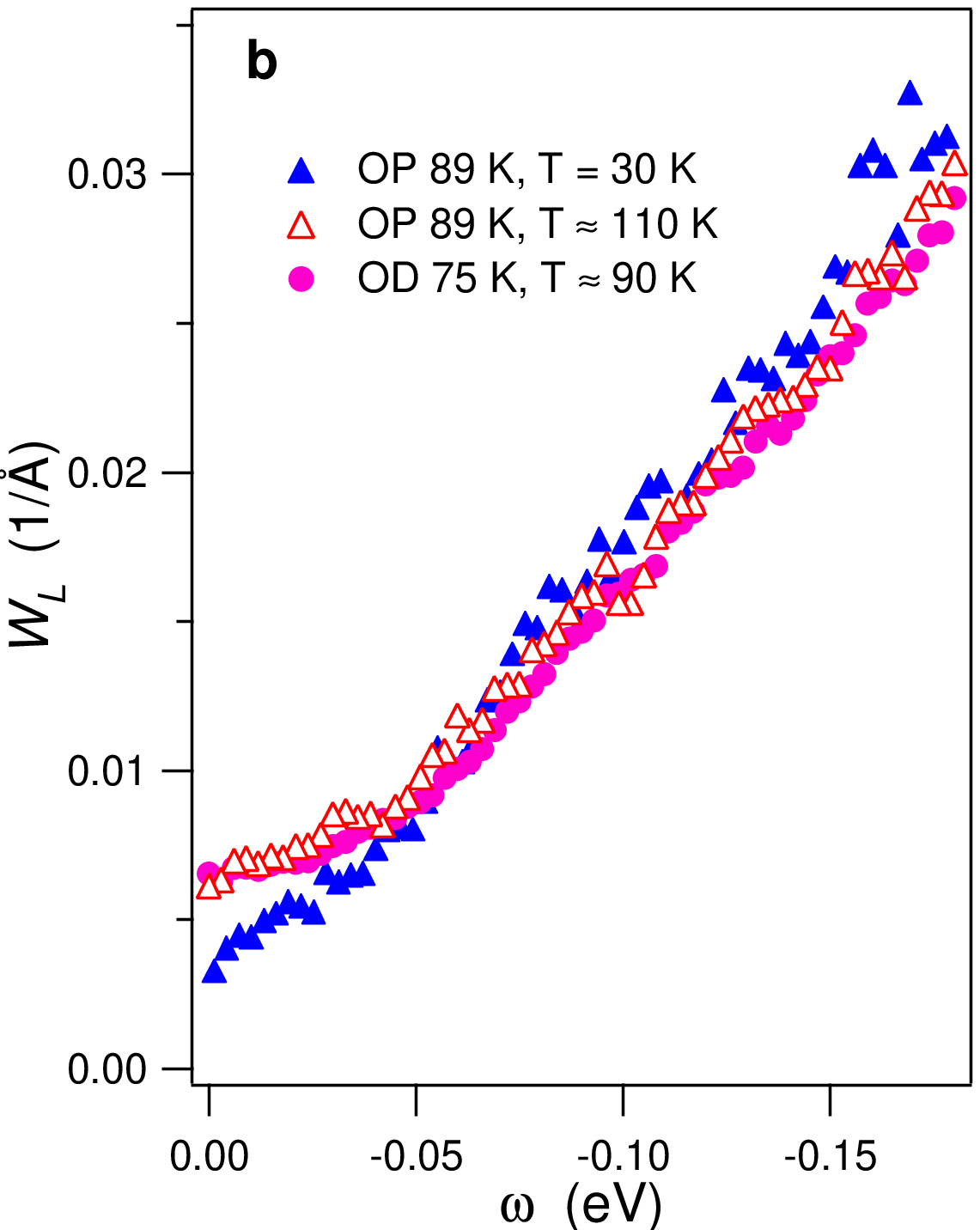}\;%
\includegraphics[width=5.4cm]{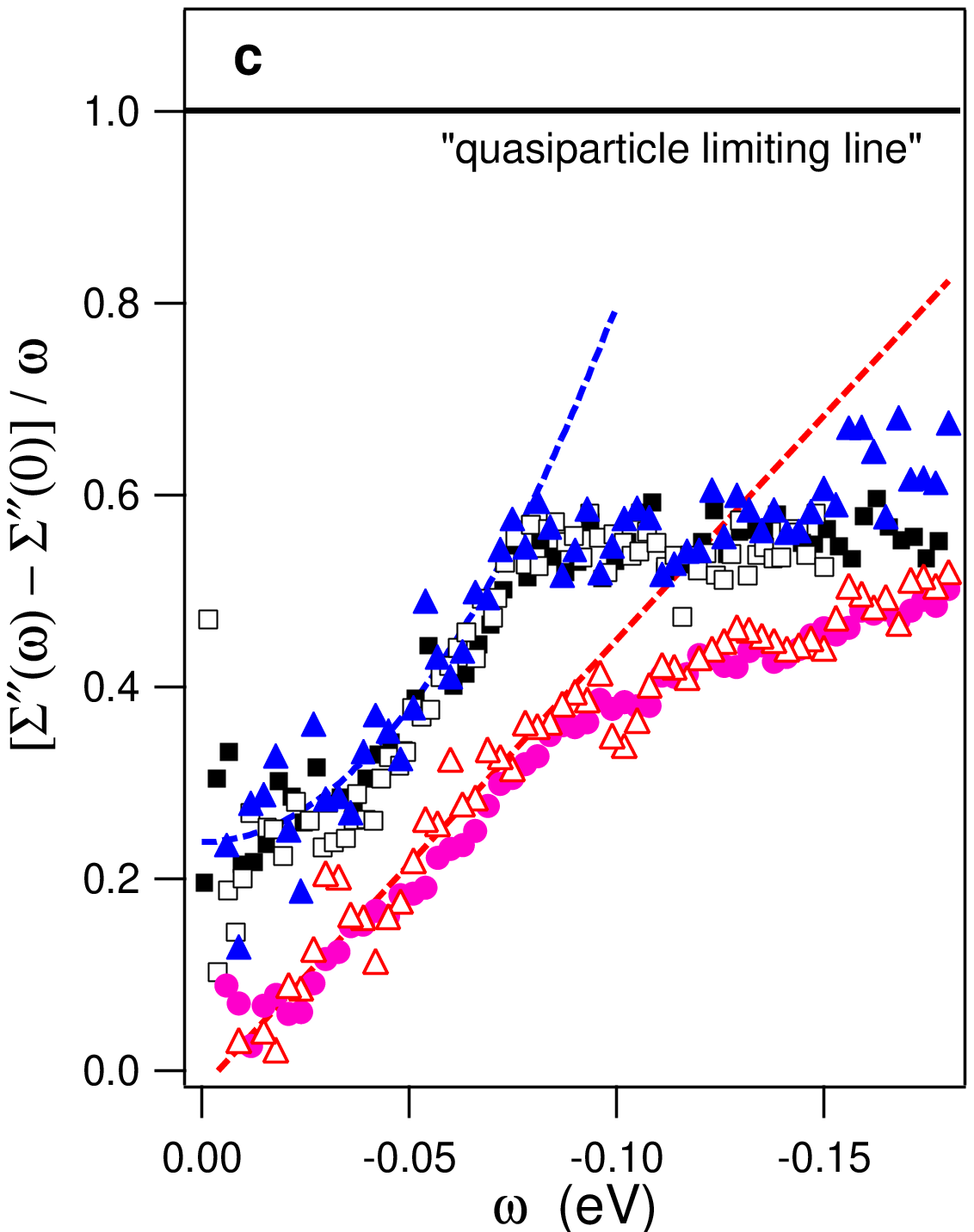}%
\caption{\label{Fig3} Energy dependence of the scattering rate. (a) Optimally doped Bi-2212 below $T_c$: MDC width (HWHM) measured with 27\;eV synchrotron radiation \cite{KordyukPRL2004} (filled squares) and the true width of the quasiparticle spectrum for the same sample purified from the resolution effect (filled triangles), to compare to the data for similar sample measured with 6 eV laser light \cite{KoralekPRL2006} (open squares). (b) Evolution of $W(\omega)$ with increasing temperature and overdoping. (c) The same data shown in a reduced dimensionless scale, dashed lines represent a parabolic fit to the low $T$ data and a linear fit to the hight $T$ data, respectively.}
\end{figure*}

Fig.\;\ref{Fig2} illustrates effectiveness of the fitting procedure for the nodal spectra analysis. The parameters of the fit are shown in Fig.\;\ref{Fig2}a as a function of energy for Bi-2212 OP89 sample measured above $T_c$. Here $W_V$ is the width of the Voigt profile, while $W_L$ and $W_G$ are the widths of its constituents, the Lorentzian and Gaussian, respectively. $W_{\mathrm{lor}}$ represents the ``traditional'' MDC width used in previous data analysis---the width of the Lorentzian from the pure Lorentzian fit. Fitting MDC's to the Voigt profile instead of to the Lorentzian introduces one additional parameter, $W_G$. In the fitting procedure we use the Voigt function \cite{Voigt} implemented in IGOR Pro (Wavemetrics Inc.) which can be approximated by a simple relation between the above mentioned parameters: 
\begin{equation}
\label{Voigt}W_V = V(W_L, W_G) = \frac{W_L}{2}+\sqrt{\frac{W_L^2}{4}+W_G^2}.
\end{equation}
Deviation of this approximation from real convolution is maximal when $W_L \approx W_G$ but is less than 1.2\%.

As it is expected for contribution of the experimental resolution, the Gaussian width is essentially $\omega$-independent in the actual range of interest, $-0.1\;\mathrm{eV} < \omega < 0\;\mathrm{eV}$: $W_G(\omega) = W_{G}^{f}$. So, the energy dependence of $W_V$ is accumulated in $W_L$, which represents now a \textit{true} width of the quasiparticle spectral function. At higher binding energies ($\omega < -0.1\;\mathrm{eV}$), $W_G(\omega)$ starts to increase that, as it follows from our simulations, can be explained by a non-linear bare dispersion. For even lower energies ($< -0.2$\;eV), due to critical lowering of the signal to noise ratio, the fit becomes unstable in distinguishing the lineshape type. Therefore, in order to reduce the experimental uncertainty we fit the $W_G(\omega)$ to $W_{G}^{f}$ on [$-0.1, 0$]\;eV energy range and define the true Lorentzian widths $W_{L}^{f}$ from Eq.\;(1): $W_V = V(W_L^{f}, W_G^{f})$. In the following, we omit ``$f$" but discuss exactly the $W_{L}^{f}(\omega)$ function as the most careful representative of the true scattering rate.

Fig.\;\ref{Fig2}b represents not usual but interesting example that demonstrates the efficiency of the Voigt fitting procedure. It shows the values $W_V(0)$, $W_L(0)$, and $W_G$ determined for the same sample at different temperatures. In this case, the APRES spectra have been measured during heating that implies a motion of the sample in respect to the beam spot due to thermal expansion of the manipulator. Evidently, this has resulted in different angular resolution in each point, that appeared as a random distribution of $W_G(T)$ while the recovered $W_L(T)$ function has found to be monotonic.

We have applied the procedure to a number of ARPES spectra taken from Bi-2212 samples of different doping levels at different temperatures. In following we discuss the energy dependence of the true scattering rate, $W_L(\omega)$, its zero energy value, $W_L(0)$, and its dependence on doping and temperature.

In respect to $W_L(\omega)$ problem, Eq.\;(\ref{Voigt}) helps to make an important remark. In a low scattering limit, when $W_L \ll W_G$, $W_V \approx W_G + W_L/2$. In other words, if, for example, $W_L(\omega) \propto \omega$, such a linear dependence cannot be camouflaged by the resolution. So, the use of the Voigt profile instead of the Lorentzian does not influence the qualitative result for the dependence of the scattering rate on binding energy \cite{KordyukPRL2004}---one can see no principal difference between the $W_L(\omega)$ and $W_{\mathrm{lor}}(\omega)$ functions in Fig.\;\ref{Fig2}a---but becomes highly important if we are interested in its careful quantitative analysis \cite{KordyukPRB2005}. 

\begin{figure}[!t]
\includegraphics[width=7.6cm]{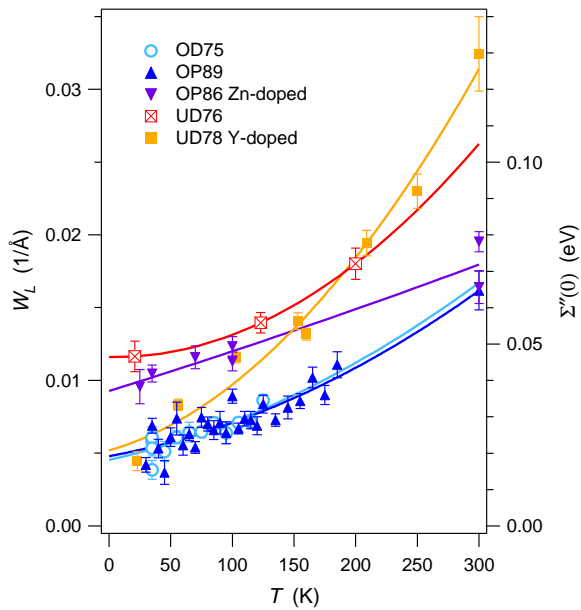}%
\caption{\label{Fig4} Temperature dependence of the scattering rate. Error bars are determined by mean square deviation of $W_G^f$, solid lines show the results of 2nd order polynomial fitting.}
\end{figure}

With Fig.\;\ref{Fig3} we discuss the energy dependence of the scattering rate. In panel (a) we compare the MDC width for OP89 sample, measured below $T_c$ with 27\;eV synchrotron radiation \cite{KordyukPRL2004}, to the data for the near-optimally doped Bi-2212, measured with 6 eV laser light \cite{KoralekPRL2006}. Besides different offsets, two datasets look very similar, [see also panel (c)]. Moreover, when the width of the quasiparticle spectrum for OP89 sample is purified from the resolution effect, it almost coincides with the laser data. This supports the conclusions about validity of the sudden approximation and that the ARPES spectra on Bi-2212 represent its bulk properties.

To reveal the asymptotic behavior of the scattering rate at low energy, in panel (c) we replot these data in form of $\sigma''(\omega) = [\Sigma''(\omega) - \Sigma''(0)]/ \omega$. To get $\Sigma''$ we just multiply $W_L$ by $v_F =$ 4\;eV\AA\;\cite{KordyukPRB2005}. One can see that, below $T_c$ at low energy, $\sigma''(\omega)$ has a non-vanishing offset and a quadratic term. In $\Sigma''(\omega)$, these terms correspond to $\sim \omega$ and $\sim \omega^3$ terms, respectively. At higher temperatures and hole doping [see panels (b) and (c)], $\Sigma''(\omega)$ becomes purely quadratic. At very low energy ($|\omega| <$ 50\;meV), both terms below $T_c$ can be explained by influence of the $d$-wave superconducting gap which causes the density of states (DOS) to become linear at low energy: the linear term is expected for the elastic forward scattering and the cubic term can be associated with inelastic spin-fluctuation scattering \cite{DahmXXX2005}. 
Also, in panel (c), we draw the ``quasiparticle limiting line'' to show how, in fact, well defined are the nodal quasiparticles in Bi-2212.

Fig.\;\ref{Fig4} shows $W_L(0,T)$ dependences for several samples. It is important to stress that the scattering rate, even when purified from resolution, does not vanish at $\omega \rightarrow 0$ and $T \rightarrow 0$. The residual, $\Sigma''_0 = v_F W_L(0,0)$, can be naturally explained by scattering on impurities. It has been shown recently \cite{DahmXXX2005} that such a high value ($\Sigma''_0 >$ 16\;meV or $W_L(0,0) > 4\times10^{-3}$\;\AA$^{-1}$) can be caused by a forward scattering on out-of-plane impurities rather than by usual unitary scattering on in-plane impurities. However, we should note, that the determined $\Sigma''_0$ values are more than 7 times lower than the values compared earlier to resistivity measurements \cite{YoshidaPhB2004}, therefore the actual discrepancy between transport and ARPES experiments is much lower.

Another conclusion that can be derived from the offset values in Fig.\;\ref{Fig4} is that $\Sigma''_0$ does not show a systematic dependence on doping level (OD75, OP89 and UD78 samples), but is rather sensitive to impurity concentration (here UD76 represents a sample, highly underdoped by annealing in He, which removes oxygen from SrO planes introducing a number of out-of-plan defects, while OP86 sample doped with 1\% of Zn gives an example of the in-plane impurity). This allows to conclude that the increase of the residual resistivity with underdoping still cannot be explained by an increase of the scattering rate but only by a decrease of the charge carrier density, which is inconsistent with ``large'' Fermi surface with area $\sim 1-x$ \cite{KordyukPRB2002}, providing therefore an indirect evidence for the phase separation in underdoped cuprates.

$T$-dependence of $\Sigma''$ deserves a separate investigation and discussion. One of the surprising results is that, within the accuracy $\sim$ 10\;meV, there is no drop of $\Sigma''$ detected at $T_c$. This can be a consequence of the pseudo-gap, which smoothes the $T$-dependence of the phase space available for scattering. Such $T$-dependence is a reason for a $\omega-T$ asymmetry of $\Sigma''(\omega,T)$, and comparison of $\Sigma''(0,T)$ and $\Sigma''(\omega,0)$ can be used to determine the phase space evolution. It is also important to understand the reason for different dependences presented in Fig.\;\ref{Fig4}. While for the most samples the variation in $\Sigma''(0,T)$ can be explained by different scattering on impurities, the much stronger $T$-dependence for the Y-doped sample needs further investigations. Answering these questions require data of much better accuracy. The described procedure eliminates the sample-related uncertainty, so, the final accuracy depends only on experimental statistics. Therefore, only time is needed to clarify these points.

The project is part of the Forschergruppe FOR538. The work in Lausanne was supported by the Swiss National Science Foundation and by the MaNEP.


\begin{thebibliography}{}
\bibitem{QP} Here we call \textquotedblleft quasiparticles" any elementary electronic excitations which can be described in terms of Dyson self-energy. The excitations with a non-zero coherent component \cite{Hufner} we will call \textquotedblleft well-defined quasiparticles".
\end{thebibliography}
\end{document}